\documentclass[fleqn,usenatbib]{mnras}

\usepackage{newtxtext,newtxmath}

\usepackage[T1]{fontenc}
\usepackage{ae,aecompl}

\usepackage{graphicx}	
\usepackage{amsmath}	
\usepackage{amssymb}	

\title[Environment and the UV Upturn]{Environmental Effects on the UV Upturn in Local Clusters of Galaxies}

\author[S. S. Ali et al.]{
Sadman S. Ali,$^{1,2}$\thanks{E-mail: s.ali@bristol.ac.uk}
Malcolm N. Bremer,$^{2}$
Steven Phillipps$^{2}$
and Roberto De Propris$^{3}$
\\
$^{1}$Subaru Telescope, NAOJ, 650 N. Aohoku Place, Hilo, HI 97620, USA\\
$^{2}$H. H. Wills Physics Laboratory, University of Bristol,
Tyndall Avenue, Bristol, BS8 1TL, United Kingdom\\
$^{3}$FINCA, University of Turku, Vesilinnantie 5, Turku, 20014, Finland\\
}

\date{Accepted XXX. Received YYY; in original form ZZZ}

\pubyear{2019}

\begin{document}
\label{firstpage}
\pagerange{\pageref{firstpage}--\pageref{lastpage}}
\maketitle

\begin{abstract}
We explore the dependence of UV upturn colours in early type
cluster galaxies on the properties of their parent clusters
(such as velocity dispersion and X-ray luminosity)
and on the positions and kinematics of galaxies within them.
We use a sample of 24 nearby clusters with highly complete
spectroscopy and optical/infrared data to select a suitable
sample of red sequence galaxies, whose FUV and NUV magnitudes
we measure from archival GALEX data. Our results show that the
UV upturn colour has no dependence on cluster properties 
and has the same range in all clusters. There is also no
dependence on the projected position within clusters or on
line-of-sight velocity. Therefore, our conclusion is that the 
UV upturn phenomenon is an intrinsic feature of cluster early
type galaxies, irrespective of their cluster environment.

\end{abstract}




\section{Introduction}

\begin{table*}
\centering
\caption{Summary of Observations}
\label{tab:1}
\begin{tabular}{lccccccc}
\hline
Cluster & RA (2000) & $\delta$ (2000) & $z$ & $\log L_X$ & $\sigma$  & Optical photometry & GALEX images \\
 & hh:mm:ss & dd:mm:ss & & 0.1--2.4 KeV & km s$^{-1}$ & & \\
\hline
\hline
Abell 930 & 10 06 54.6 & $-05$ 37 40 & 0.0549 & 35.78 & 907 & 
SDSS, PS1 & AIS 315 \\
Abell 954 & 10 13 44.8 & $-00$ 06 31 & 0.0932 & 37.91 & 832 & 
SDSS, PS1 & MIS DR1 \\
Abell 957 & 10 13 40.3 & $-00$ 54 52 & 0.0436 & 36.60 & 640 &
SDSS, PS1 & MIS DR1, DR2 \\
Abell 1139 & 10 54 04.3 & +01 29 56 & 0.0398 & 37.33 & 503 & SDSS, PS1 & MIS DR2 \\
Abell 1189 & 11 11 04.0 & +01 07 42 & 0.0962 & 36.19 & 814 &
SDSS, PS1 & MIS DR1 \\
Abell 1236 & 11 22 44.9 & +00 27 44 & 0.102 & 36.42 & 589 & 
SDSS, PS1 & MIS DR1 \\
Abell 1238 & 11 22 58.0 & +01 05 32 & 0.0733 & 36.49 & 586 & SDSS, PS1 & MIS WZN11 \\
Abell 1364 & 11 43 39.6 & $-01$ 45 39 & 0.106 & 35.85 & 600 &
SDSS, PS1 & GI5--GAMA12, MISGCSN \\
Abell 1620 & 12 49 46.1 & $-01$ 35 20 & 0.0821 & 34.48 & 1095 & SDSS, PS1 & MIS DR1 \\
Abell 1663 & 13 02 50.7 & $-02$ 30 22 & 0.0843 & 37.00 & 884 & SDSS, PS1 & MIS DR1 \\
Abell 1692 & 13 12 16.0 & $-00$ 55 55 & 0.0842 & 36.75 & 686 & 
SDSS, PS1 & GI4 \\
Abell 1750 & 13 30 49.9 & $-01$ 52 22 & 0.0852 & 37.50 & 981 & 
SDSS, PS1 & MISGCSN \\
Abell 2660 & 23 45 18.0 & $-25$ 58 20 & 0.0525 & 35.70 & 845 & PS1 & AIS 279 \\
Abell 2734 & 00 11 20.7 & $-28$ 51 18 & 0.0618 & 37.41 & 780 & 
PS1 & GI1 \\
Abell 2780 & 00 29 17.1 & $-29$ 23 25 & 0.0988 & ... & 782 & PS1 & AIS 280 \\
Abell 3094 & 03 11 25.0 & $-26$ 53 59 & 0.0677 & 36.76 & 774 & 
PS1 & AIS 405 \\
Abell 3880 & 22 27 52.4 & $-30$ 34 12 & 0.0548 & 37.27 & 855 & 
PS1 & MIS2DFSGP \\
Abell 4013 & 23 31 51.8 & $-35$ 16 26 & 0.0500 & ... & 904 & SuperCosmos & AIS 394 \\
Abell 4053 & 23 54 46.7 & $-27$ 40 18 & 0.0720 & ... & 994 & PS1 & AIS 280 \\
Abell S0003 & 00 03 09.5 & $-27$ 53 18 & 0.0644 & ... & 833 & PS1 & AIS 280 \\
Abell S0084 & 00 09 24.0 & $-29$ 31 28 & 0.108 & 37.41 & 807 &
PS1 & MIS2DFSGP \\
Abell S0166 & 01 34 23.4 & $-31$ 35 39 & 0.0697 & ... & 511 &
SuperCosmos & AIS 407 \\
Abell S1043 & 22 36 26.8 & $-24$ 20 26 & 0.0340 & ... & 1345 & PS1 & MIS2DFSGP \\
\hline
\end{tabular}
\end{table*}

The Ultraviolet Upturn or Excess is an unexpected rise in flux in the spectral energy distributions of early-type galaxies shortwards of 2500 \AA. It  appears to be a nearly ubiquitous property of spheroids and bulge-dominated galaxies (see, e.g., \citealt{Yi2008,Yi2010} for a recent review) although their generally old, metal-rich and quiescent stellar populations (e.g., \citealt{Thomas2005,Thomas2010})
should contain no sources capable of providing significant flux below the 4000 \AA\ break. While many candidates have been proposed, the source population is generally agreed to consist
of hot horizontal branch (HB) stars
\citep{Greggio1990,Brown1998a} and it now appears most 
likely that this population is helium-rich and formed {\it in situ} at high redshift 
\citep{Ali2018a,Ali2018b,Ali2018c}. While such stars are now known to exist in globular
clusters in our Galaxy (e.g., see \citealt{Norris2004,Piotto2005,Piotto2007}) and likely
elsewhere \citep{Kaviraj2007a,Mieske2008,Peacock2017}, the origin of such high helium 
abundances is unclear\footnote{It must be noted that this hypothesis was originally presented 
by \cite{Hartwick1968} and \cite{Faulkner1972}, who also commented on the `unpalatability' 
of the proposed solution.}. It has been suggested that the effect may depend on the environment: 
stratification of helium in the centres of clusters might create populations of galaxies with 
high helium abundances \citep{Peng2009}. On the other hand, it is also possible that the helium 
enrichment depends on the details of early star formation, as in globular clusters. Differences 
in the originating mechanisms or timescales may be reflected in different evolutionary histories 
for galaxies in clusters and the field. \cite{Atlee2009} finds a slow decrease in the strength 
of the upturn for a sample of bright field elliptical galaxies at $0 < z < 0.65$ compared
to the nearly constant colour for very bright ellipticals in the works of \cite{Brown1998b,Brown2000,Brown2003}
and brightest cluster galaxies \citep{Loubser2011,Boissier2018}. In our work on much larger samples 
of cluster early types with luminosities down to $L^*$,  we also see no evolution at $z < 0.55$ 
\citep{Ali2018a,Ali2018b,Ali2018c} but then detect a rapid reddening in the UV colour at $z=0.7$; 
this may be consistent with the last data point in \cite{Atlee2009}, despite the large errors
and small number statistics. \cite{LeCras2016} also find evidence for evolution in the UV upturn 
at $z > 0.6$, albeit from stacked spectra and using spectroscopic indices sensitive to the HB population, for a sample of very luminous BOSS 
galaxies (as opposed to the UV photometry used by other studies).

\cite{Ali2018a} compared the UV upturn colours for galaxies in the Coma, Fornax and Perseus clusters, 
and found no evidence that the UV upturn was affected by the environment, a conclusion also reached by 
\cite{Smith2012} when comparing galaxies in the Coma and Virgo clusters. Similarly, the UV upturn colour 
of brightest cluster galaxies in the samples of \cite{Loubser2011} and \cite{Boissier2018} did not appear 
to depend strongly on cluster properties. However, it would be interesting to extend this to several cluster 
environments and consider the eventual dependence on position within the cluster and on kinematics. For example, 
if residual star formation contributes to the UV flux, as argued by \cite{rich2005,yi2005,salim2010}
then one would expect to observe a dependence on cluster properties (e.g., X-ray luminosity if ram stripping is
important) and/or on position within the cluster or kinematics (as a proxy for orbits that avoid or pass through 
the cluster core for instance). Although the effect may be stochastic, it should emerge at some level in the 
ensemble of several clusters studied here.

Furthermore, the Helium sedimentation model of \cite{Peng2009} predicted that the strength of the UV upturn 
should be stronger in a) larger mass clusters; b) clusters with cooling flows and; c) in dynamically relaxed 
clusters. The model also predicts that UV upturn galaxies should be more prevalent in cluster cores. However, 
these key environmental dependencies of the upturn strength were proven to not hold true by the studies of 
\cite{Donahue2010} and \cite{Loubser2011}, who found no correlation between the $FUV-NUV$ colour (a measure 
of the upturn strength) in low redshift cluster galaxies and the aforementioned parameters (see also \citealt{Yi2011}).


In this paper we exploit a highly complete sample of galaxies in 24 nearby clusters to derive the dependence 
of the UV upturn color on cluster properties and on the local cluster environment. We describe the sample and 
photometry in the next section and present the results in section 3. Discussion of our findings and conclusions 
are shown in section 4. We assume the conventional cosmological parameters from the latest Planck datasets.

\section{Database}

The sample we studied here consists of galaxies in 24 clusters from the sample of \cite{DePropris2017}. These clusters span a wide range of properties in terms of velocity dispersion, 
X-ray luminosity and  Bautz-Morgan type and therefore allow us to consider how the UV upturn colour is affected by cluster
properties over a large range in environmental density, hot 
gas content and dynamical indicators (e.g.,  relaxed vs. merging clusters). In particular, we can explore the regime
between massive systems, such  as Coma or Perseus, and poor clusters and rich groups (similar, in some respect to Virgo or Fornax).

Each cluster was observed with the CTIO 4m telescope in the $K_s$ band, to produce a deep (300s exposure) 
image of the entire cluster across its Abell radius.The $K_s$ luminosity is found, empirically, to be an 
excellent proxy for stellar mass \citep{Gavazzi1996,Bell2003,Kettletty2018} and this therefore provides a 
stellar mass selected sample of galaxies in nearby clusters (mean redshift of 0.075). Galaxies in these 
clusters down to at least the level of the measured $K^*+2$ were identified as spectroscopic members (with 
typical completeness of about 80\% even in the faintest luminosity bin considered) from existing spectroscopic 
data. This also provides information on the kinematics of these galaxies within each cluster.

For all galaxies we derived optical colours ($g-r$ or $B_J-R_F$) from data in the PanStarrs1 survey \citep{Chambers2016,
Flewelling2016,Magnier2016} or (for objects below --30$^{\circ}$) in the SuperCosmos survey \citep{Hambly2001a,Hambly2001b}. 

We then matched all confirmed spectroscopic members to FUV and NUV data in the GALEX database \citep{Morrissey2007}.
We used a $4.5''$ matching radius and only selected objects with S/N of at least 5 in the GALEX NUV photometry. These 
come from a combination of AIS and MIS imaging. Exposure times range from a few ks in MIS data to a few hundred seconds 
in AIS. For blue cloud galaxies in our clusters, 93\% have a NUV detection and 76\% have a FUV detection, whereas 72\%
of red sequence galaxies have a NUV detection and 40\% have a FUV detection. We can assume that objects with no NUV 
detection have red colours, given the high detection fraction for blue cloud galaxies.

In Table~\ref{tab:1} we show the main properties of each cluster used and the relevant information on the sources of
photometry (see also \citealt{DePropris2018}). All data were corrected for Galactic extinction using the latest values from 
\cite{Schlafly2011} and extrapolating to the FUV and NUV bands with a Milky Way extinction law (e.g., \citealt{Calzetti1994}).
Optical and infrared data were corrected for extinction with the same procedures. Colours were also $k$-corrected using the derived UV spectral energy distribution (SED) from \cite{Ali2018a}. This uses a standard model from \cite{Conroy2009} for the optical and a 16000K blackbody for the contribution due to the vacuum UV light, a combination that fits the observed spectral energy distributions of Coma cluster galaxies across the whole range from 1000 to 10000 \AA.

\section{Results}

\subsection{Identification of quiescent galaxies}

In our previous papers \citep{DePropris2017,DePropris2018} we
have identified galaxies on the red sequence and blue cloud by
fitting a straight line to the colour-magnitude relation (in each cluster, separately) using a minimum absolute deviation
method that discriminates against outliers (see \citealt{DePropris2017,DePropris2018} for details). The 1$\sigma$ scatter in $g-r$ for red sequence galaxies
was measured to be 0.05 mag. and we therefore selected galaxies as belonging to the red sequence if they lie within $\pm 0.15$ 
mag. of the best-fitting straight line to the colour-magnitude relation. Bluer cluster members are assigned to the blue cloud.

\begin{figure}
    \centering
    \includegraphics[width=\columnwidth]{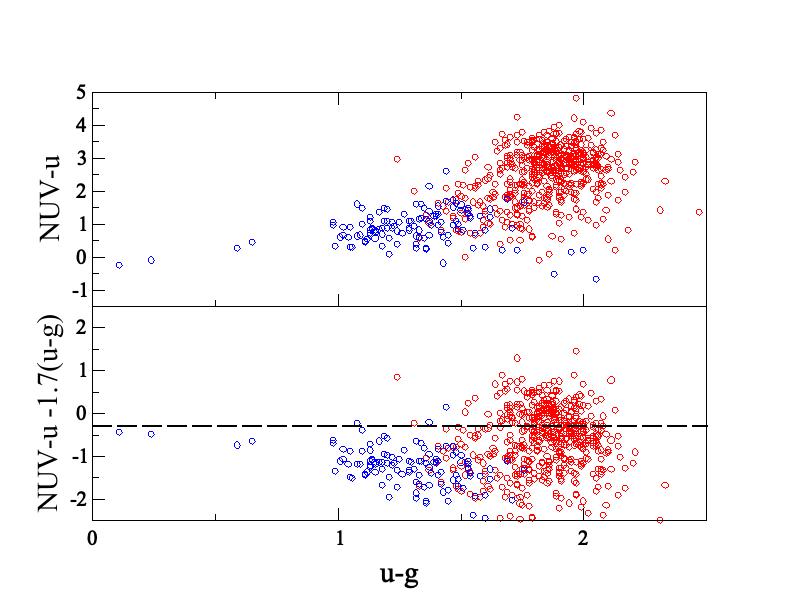}
    \includegraphics[width=\columnwidth]{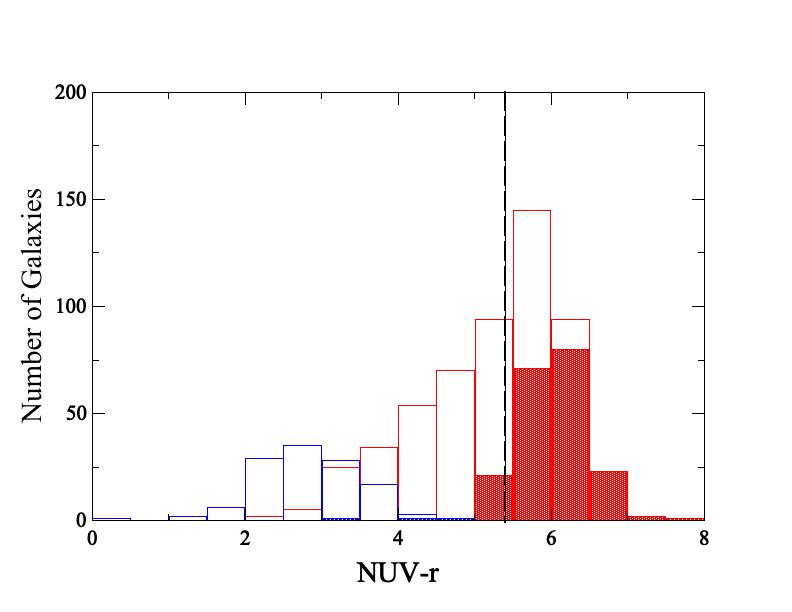}
    \caption{Top: The colour-colour plot ($NUV-u$ vs. $u-g$
    for galaxies in 11 of our clusters where SDSS $u$ data are
    available. Middle: We define a vector $y=NUV-r - 1.7 (u-g)$ to remove the linear trend observed in the top panel between $NUV-u$ and $u-r$ and plot this vector vs. 
    $u-g$ to show
    that star-forming galaxies generally have $y < -0.3$. 
    Bottom: We plot the numbers of galaxies in the initial
    sample and objects with $y < -0.3$ (open histograms)
    and $y > -0.3$ (filled histograms) as a function of
    $NUV-r$ colour. Red and orange histograms are for red
    sequence galaxies (after and before selection using the
    $y$ vector) and blue histograms are for blue cloud galaxies. We also show a line at $NUV-r=5.4$. This excludes
    nearly all star-forming galaxies from the sample.}
    \label{fig:0}
\end{figure}

Red sequence galaxies (selected in the optical) consist of
truly quiescent objects (whose UV light is produced by the UV
upturn) and galaxies with residual star formation (sometimes
called the  `green valley' -- e.g., \citealt{rich2005,
salim2010}). Several studies have adopted cuts in
$NUV - optical$ colours to separate quiescent galaxies from
those with residual star formation. \cite{Kaviraj2007} place
their selection at $NUV-r > 5.0$, while \cite{Crossett2014}
adopt a somewhat more stringent limit $NUV-r > 5.4$ to their
sample of $z < 0.1$ clusters. Here we choose this 
latter definition. 

\begin{figure}
    \centering
    \includegraphics[width=\columnwidth]{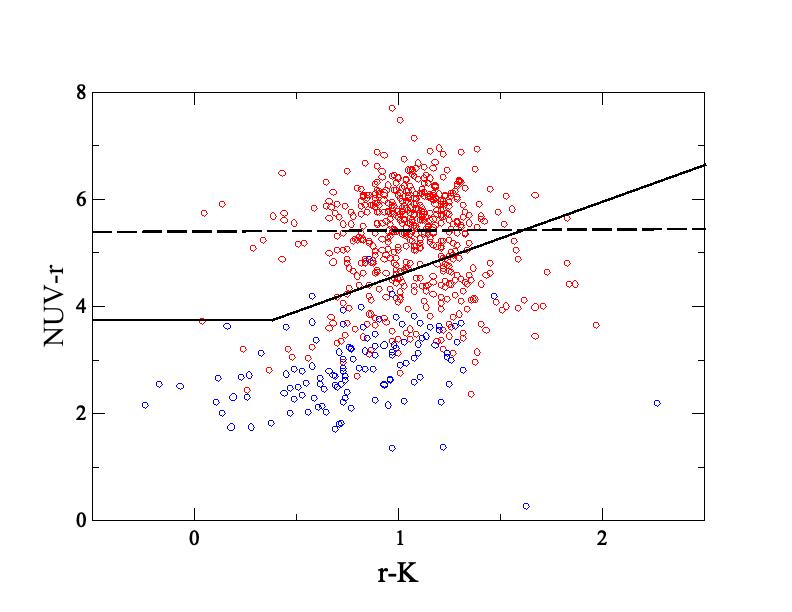}
    \caption{The colour-colour plot ($NUV-r$ vs. $r-K$
    for galaxies in our clusters (red dots for red sequence
    galaxies and blue dots for blue cloud ones). We plot
    (thick black line) the NRK vector ($NUV-r=3.75$ and
    $NUV-r=1.37(r-K)+3.2$ defined by Arnouts et al. (2013)
    to separate star-forming and quiescent galaxies. The
    dashed line shows a colour cut $NUV-r > 5.4$: this always
    lies above the NRK vector and therefore selects quiescent
    galaxies effectively.}
    \label{fig:00}
\end{figure}

Below, we show how this selection is justified from our data. We plot $NUV-r$ vs. $u-g$ for all galaxies (this is only possible for those with SDSS data, as indicated in Table~\ref{tab:1}) in Fig.~~ \ref{fig:0} (top panel). 
Here red dots are red sequence galaxies and blue dots
are galaxies in the blue cloud (as defined above). These 
latter objects follow an  approximate straight line  in this
colour-colour plane (i.e. the  so-called star-forming main
sequence -- e.g., \citealt{Speagle2014}). We fit this with 
a straight line and remove this trend from all objects, 
resulting in the middle panel of Fig.~\ref{fig:0}. By 
choosing a limit $y=(NUV-u) - 1.7(u-g) > -0.3$ we can 
exclude the vast majority of objects with star formation from
the sample. However, note that we cannot apply this 
selection to all our clusters, as those in the South have
no $u$ data (this is not provided by PanStarrs1). In 
the bottom panel of this figure we plot the number counts 
of galaxies as a function of $NUV-r$ color 
before and after the above selection. Almost all star-forming
galaxies have $NUV-r < 5.0$ and our $NUV-r > 5.4$
criterion appears to isolate a nearly pure sample of 
quiescent systems (see also Phillipps et al. 2019, in 
preparation for a detailed discussion).

\cite{Arnouts2013} estimate the star formation rate in galaxies
by fitting the full spectral energy distributions and then
derive a vector in the $NUV-r$ vs. $r-K$ plane that separates
star-forming and quiescent galaxies (cf. Phillipps et al.
2019, where we use MAGPHYS derived star formation rates to the same purpose). They confirm this approach from a
morphologically selected sample in \cite{Moutard2016a,
Moutard2016b}. We plot this colour-colour plot in
Fig.~\ref{fig:00} with the same colour scheme as in
Fig.~\ref{fig:0}. As we see a colour cut at $NUV-r > 5.4$
lies well above their NRK vector. We therefore conclude 
that selecting galaxies with $NUV-r > 5.4$ produces a 
sample consisting largely of passive galaxies whose
UV light is dominated by upturn sources, whereas the
original $NUV-r > 5.0$ cut by \cite{Kaviraj2007} may 
still include a small fraction of objects with residual star formation.
\begin{figure*}
	\includegraphics[width=\columnwidth]{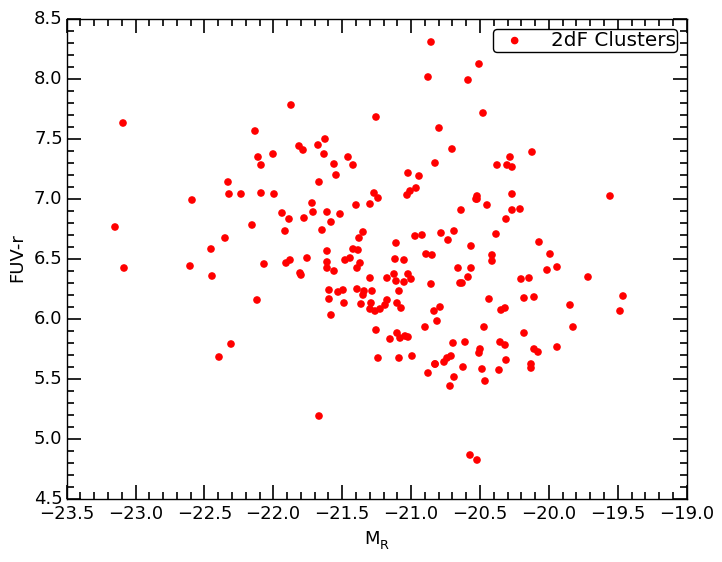}
		\includegraphics[width=\columnwidth]{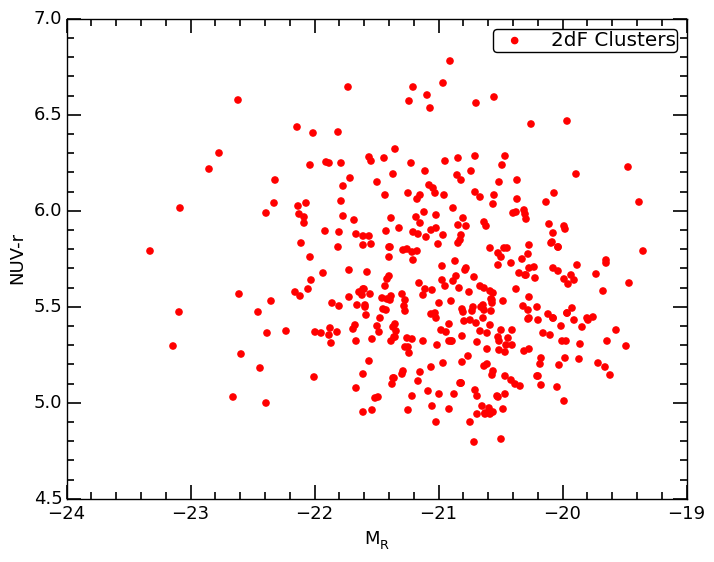}
    \caption{UV colours (FUV--r on the left and NUV--r on the right) plotted vs $M_r$ for all galaxies in the 2dF cluster
    sample of De Propris (2017). Galaxies with $NUV-r > 5.4$ are convincingly passive systems, while those with $5.0 < NUV-r < 5.4$ may still contain some residual star formation. Opposite to the $\sim 1.5$ -- $2$ mag. scatter in these colours, the scatter in $g-r$ is
    about 0.05 mag.}
    \label{fig:1}
\end{figure*}

\begin{figure*}
	\includegraphics[width=\columnwidth]{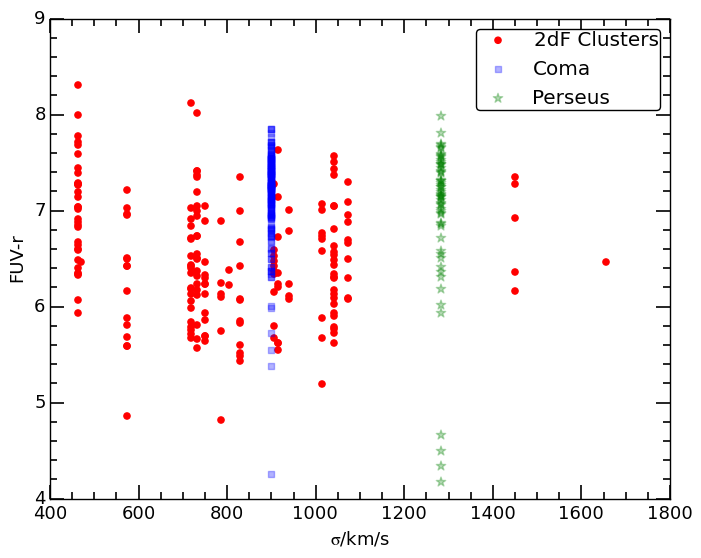}
		\includegraphics[width=\columnwidth]{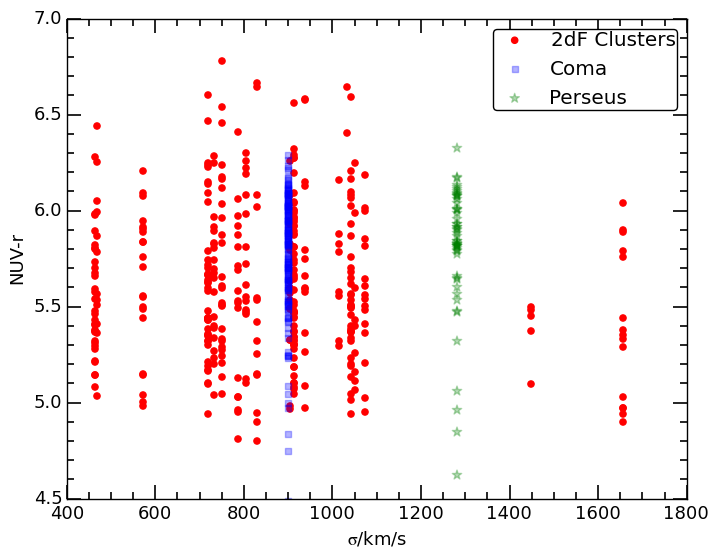}\\
	\includegraphics[width=\columnwidth]{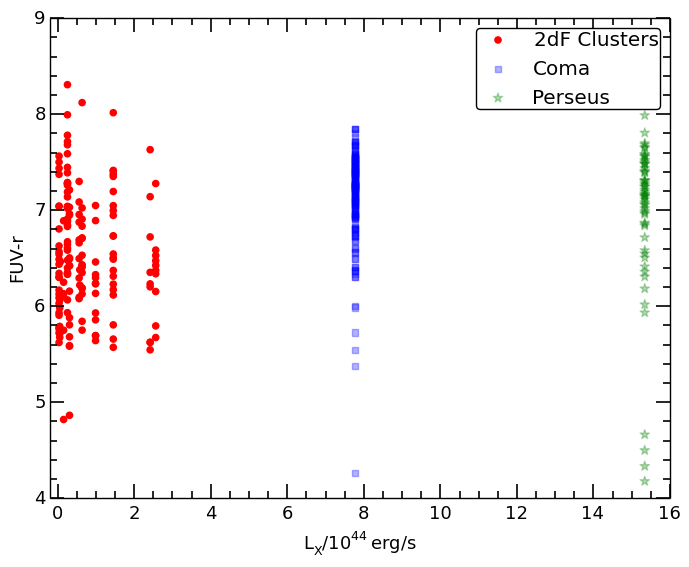}
		\includegraphics[width=\columnwidth]{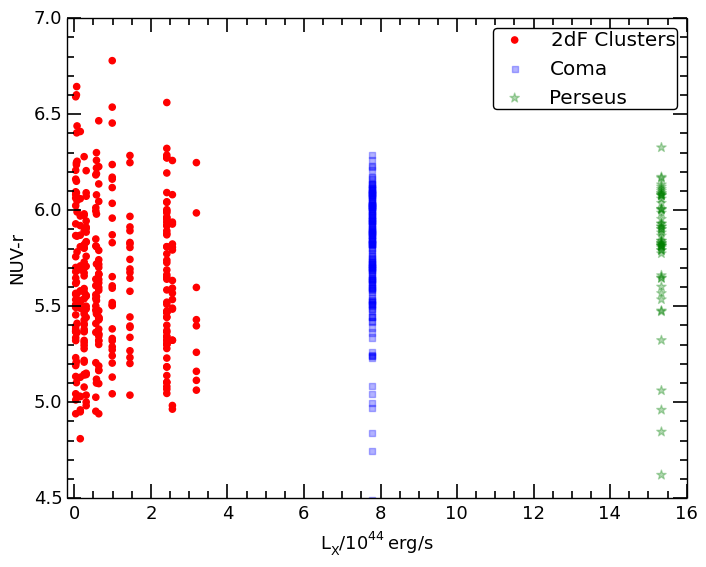}		
    \caption{Top: UV colours (FUV--V on the left and NUV--V on the right) plotted vs cluster velocity dispersion for all galaxies in the 2dF cluster sample of De Propris (2017).
    Bottom: UV colours (FUV--r on the left and NUV--r on the right) plotted vs cluster X-ray luminosity in the ROSAT soft band (0.5-2 KeV) for all galaxies in the 2dF cluster
    sample of De Propris (2017). We also show data from the literature (Ali et al. 2018a)
    for Coma and Perseus.}
  \label{fig:2}
\end{figure*}

\subsection{UV colors of quiescent galaxies}

Fig.~\ref{fig:1} shows the derived $FUV-r$ and $NUV-r$ colours
for cluster red sequence galaxies plotted against $M_r$. The 
$FUV-r$ colours mostly range between 5.5 and 7.5 mag. This
observed $\sim 2$ magnitude range in colour is similar to that
of Coma \citep{Smith2012} and Perseus \citep{Ali2018a}, as well as that observed in  Virgo red sequence galaxies
\citep{Boselli2005}. The $NUV-r$ colour varies
between 5 and 6.5 mag., and once again, the $\sim 1.5$ magnitude range observed is typical of the aforementioned low redshift cluster galaxies. This can be compared with the well-known small scatter of optical-infrared colour-magnitude relations for red sequences in clusters (e.g., \citealt{Valentinuzzi2011}). In our sample, the intrinsic
scatter is $g-r$ for red sequence galaxies is 0.05 mag. \citep{DePropris2017}, i.e., nearly two orders of magnitudes
smaller than in $FUV-r$ or $NUV-r$. This consistent range in $FUV-r$ and $NUV-r$
observed between all low redshift cluster galaxies indicates
that the upturn is a universal feature among all such old,
passively evolving systems and also that the environment must
not have any significant influence on this phenomenon, since the
clusters studied here have a wide variety of properties.

We further explore this by plotting $FUV-r$ and $NUV-r$ colours
for galaxies in clusters vs. the cluster velocity dispersion (a
broad proxy of the cluster mass) and vs. the X-ray luminosity in
the ROSAT soft band (\citealt{Truemper1993}; taken from the BAX catalogue - \citealt{Sadat2004}), which is a measure of the gas
density in each cluster, in Fig.~\ref{fig:2}. We also plot
data for Coma and Perseus for comparison. It is clear that 
there is no obvious trend in these colours with cluster
properties, across a wide range of cluster masses. As we expect
that velocity dispersion and X-ray luminosity would affect any
residual star formation, the lack of correlation with these
parameters further argues that the UV upturn properties were 
established at early times, prior to the epochs at which
galaxies first felt the effects of the cluster environment.

\begin{figure*}
	\includegraphics[width=\columnwidth]{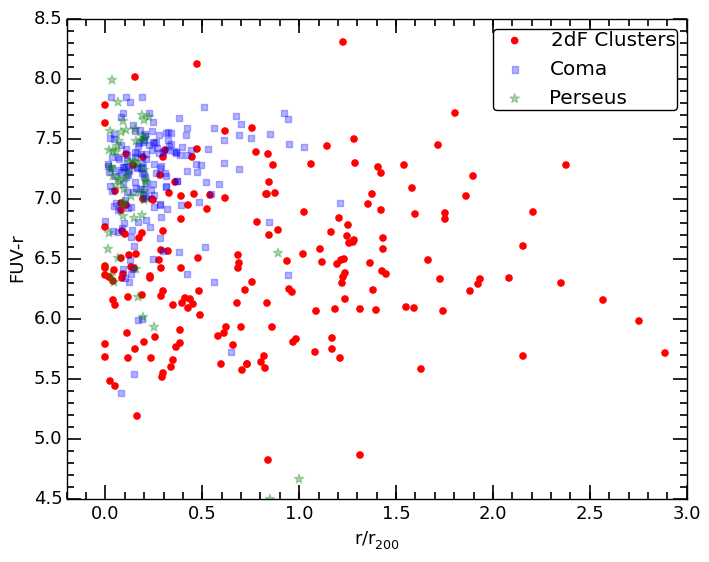}
		\includegraphics[width=\columnwidth]{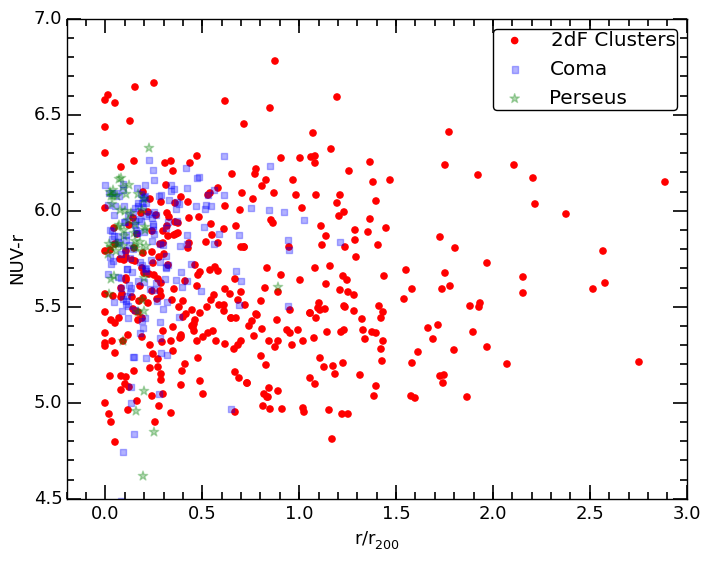}\\
   	\includegraphics[width=\columnwidth]{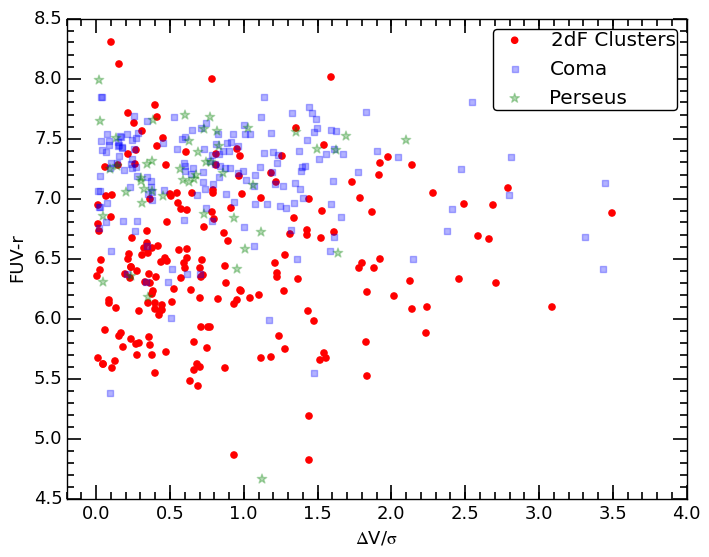}
		\includegraphics[width=\columnwidth]{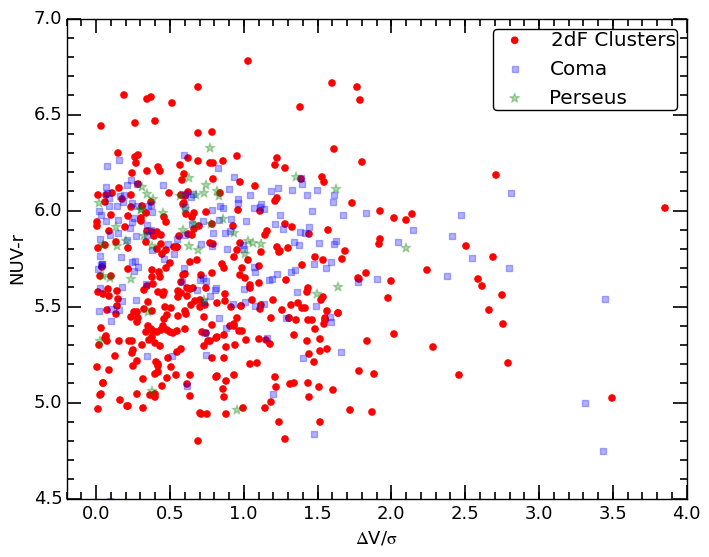}
    \caption{Top: UV colours (FUV--r on the left and NUV--r on
    the right) plotted vs $R/R_{200}$ (projected) in the 2dF
    cluster sample of De Propris (2017). Bottom: UV colours (FUV--r on the left and NUV--r on the right) plotted vs $\Delta V/ \sigma$. We also show data for Coma
        and Perseus as in Fig.~2.}
    \label{fig:4}
\end{figure*}

We next consider whether the UV upturn colour is affected by
position within the cluster or by kinematics. In Fig.~\ref{fig:4} we plot $FUV-r$ and $NUV-r$ vs $R/R_{200}$,
where $R_{200}$ is calculated from \cite{Carlberg1997} using 
data from \cite{DePropris2017}. Objects projected closer to the
centre are likely to have been in the cluster core longer \citep{Smith2012b}: these are mainly classical ellipticals
in the inner 0.3 $R/R_{200}$. The colours appear identical
irrespective of projected position. Neither do we see any
dependence of these colours on $\Delta V / \sigma$
(Fig.~\ref{fig:4}), where $\Delta V$ is the difference between
the velocity of each galaxy and the mean velocity of its parent
cluster and $\sigma$ is the cluster velocity dispersion (thus objects with higher $\Delta V / \sigma$ are likely to be less
virialised and to move on more radial orbits, and therefore
to be relative newcomers to the cluster environment and to be
more affected by tidal and ram stripping processes). Therefore,
these provide further evidence against the existence of 
strong environmental effects, especially those that would
affect continuing star formation.

\begin{figure*}
	\includegraphics[width=\columnwidth]{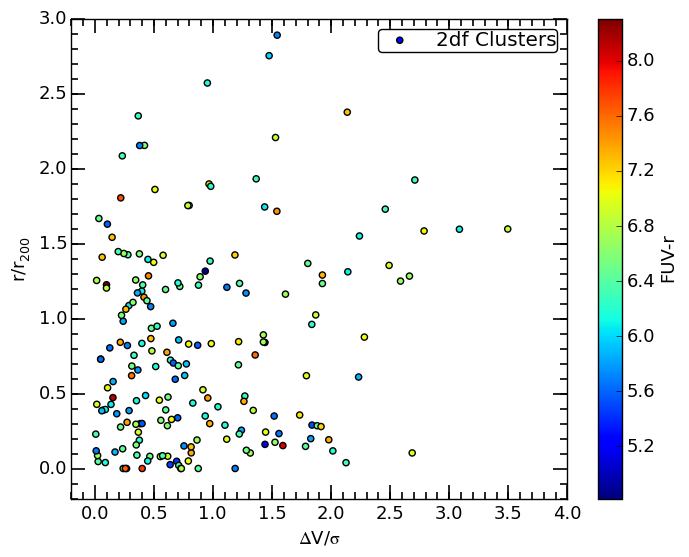}
		\includegraphics[width=\columnwidth]{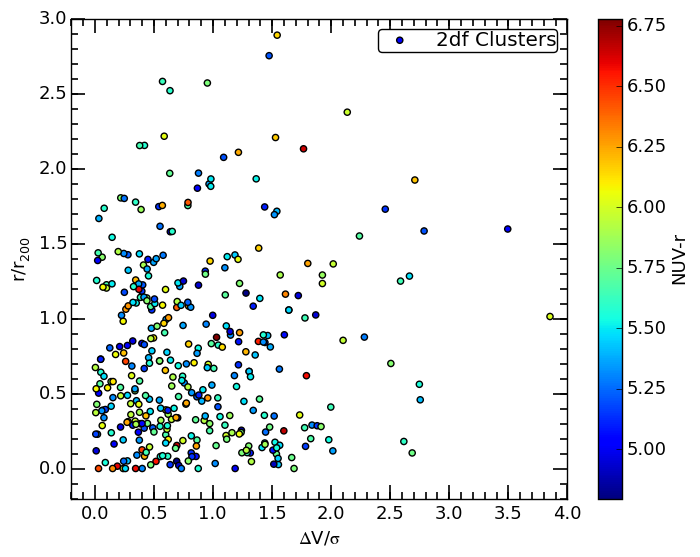}
    \caption{Caustic plots for all galaxies colour coded according to their FUV--r or NUV--r colour.}
    \label{fig:6}
\end{figure*}

Finally we plot $\Delta V / \sigma$ vs. $R/R_{200}$ for galaxies
colour-coded according to their $FUV-r$ and $NUV-r$ colours, in
Fig.~\ref{fig:6}. These caustic plots are related to the orbits
of galaxies within clusters and we observe no significant
dependence on UV upturn colour. This again suggests that the
UV upturn is unrelated to the cluster environment.

\section{Discussion}

As with Coma and Perseus in our previous paper \citep{Ali2018a},
the results obtained here can be best interpreted with the
presence of a He-enhanced sub-population of hot HB stars 
giving rise to the upturn, superimposed on top of the majority
`red and dead' population that makes up ETGs. The range in
colours can then be explained by varying amounts of
He-enhancement. Using the YEPS Stellar Population Synthesis
models of \cite{Chung2017} that incorporate He-enhancement, 
$Y \geq 0.39$ is predicted  (assuming solar metallicity and
$z_f=4$) to account for the full range in $FUV-r$ and $NUV-r$
colours seen in all of these clusters below $z=0.1$ (note that metallicity, unless highly subsolar, which is not the case for these galaxies, does not strongly affect the HB morphology and therefore the UV colours for passive galaxies, so the observed range in colours cannot be accounted for my metal abundance).
Consistent with the observations of \cite{Ali2018c}, these
models predict that the UV upturn colour is nearly constant 
to $z \approx 0.6$ (depending on $Y$ and the epoch at which 
the stars were formed) and then evolves rapidly to the red.
This is indeed observed in a $z=0.7$ cluster
\citep{Ali2018c}. It is also interesting that the
bluest UV upturn colours tend to occur among the most massive
galaxies, which also tend to be older and more metal rich
\citep{Smith2012}. These galaxies also have resided in the
cluster core for longer times \citep{Smith2012b}. Similarly,
\cite{Ali2018a} find that the more massive galaxies have a
hotter and more populated HB as well. This may be due to a
greater degree of helium enrichment but also to the effect of
larger ages in thinning the remaining stellar atmosphere.

Under dynamical equilibrium the velocity dispersion and X-ray
luminosity of a cluster directly correlate with its mass. Fig.
~\ref{fig:2} demonstrates that the upturn 
strength has no correlation with either velocity dispersion or
X-ray luminosity, and by proxy the mass of the cluster. As
such, clusters of all sizes can have a component of upturn in
their early-type population.

Within galaxy clusters, star-formation is quenched very
strongly in the centre due to processes such as ram-pressure
stripping, harassment and strangulation. The rate of
star-formation in cluster galaxies thus tends to increase 
with increasing cluster-centric distance
\citep[e.g.,][]{Dressler1997}. If star-formation was the 
{\it sole} driving mechanism behind the UV emission in the
galaxies in this sample, as argued by Yi et al. (2005), one
would  expect to see the $FUV-r$ and $NUV-r$ colours become
bluer with increasing radius, at least in an ensemble. 
However as seen in Fig.~\ref{fig:4} the UV-optical colours 
show no correlation with cluster-centric distance. This
reinforces the idea that the UV emission in these galaxies 
is indeed from an old, hot HB subpopulation and not from any
residual star-formation. Similarly, there appears to be no
correlation between the line-of-sight velocity of galaxies 
and the upturn strength, in Fig.~\ref{fig:4}. Therefore, 
the influence of the cluster's gravitational potential - 
the main driving force behind the velocities - does not 
affect the strength of the upturn. We can see from
Fig.~\ref{fig:6} that the majority of cluster members
are centred around the core of the cluster ($R/R_{200} < 1$,
$\Delta V/ \sigma < 2$), yet these central galaxies show the
full range of $FUV-r$ and $NUV-r$ colours as seen from the
entire population of galaxies. This suggests that there is 
no gradient in UV upturn colour with either cluster-centric
distance or line-of-sight velocity. 

These results are more in line with a He-enhanced HB origin 
for the UV upturn, in which case the cluster environment 
should have little effect on the emergence and prevalence of 
the upturn. The observed lack of environmental dependence
with position within the cluster implies that the extra 
helium cannot come from stratification as in the model of
\cite{Peng2009}. In a hierarchical model of structure
formation, galaxies form first at $2 < z < 6$ and the star
formation in ETGs rapidly comes to an end by $z \sim 2$
\citep{Jorgensen2017} . Galaxies then accrete in highly
over-dense regions of the universe to form clusters at $z \sim
1.5$ \citep[e.g.][]{Wen2011}. Studies have shown that cluster
red sequences are already established between
$1 < z < 2$ \citep{Newman2014}. This indicates that 
the majority of star-formation in ETGs was
completed and a mostly passively evolving stellar population,
that is observed at present, was already established before
these galaxies became part of clusters, or shortly thereafter.
While some field ETGs show evidence of residual star formation
where at least part of their stellar populations is formed
more recently (e.g., \citealt{Jeong2007,Davis2013}), this is
unlikely to be the case for cluster ETGs: our UV SEDs for 
ETGs in Coma \citep{Ali2018a} and Abell 1689 \citep{Ali2018b},
for instance, are inconsistent with a contribution from star formation to the UV light.

This is particularly important as the oldest stars in ETGs, 
ones that formed at $z \geq 4$, before the galaxies
became part of clusters \citep{Guarneri2018} are the ones 
that would have the necessary time required to evolve
from the main sequence on to the red giant branch, and 
then eventually to the horizontal branch, where they become UV-bright given sufficient He-enhancement. Since the cluster
environment particularly works to quench the star-formation
within galaxies, the main sequence population in ETGs, which 
is already red and passively evolving, is largely unaffected.
Hence, the upturn develops intrinsically within these galaxies
irrespective of the cluster environment. This also indicates
that the large He-enhancement that leads to the eventual UV
upturn in a sub-population of the main sequence in ETGs must
also occur intrinsically within the galaxies and at very early
times. 

\section*{Acknowledgements}

The Pan-STARRS1 Surveys (PS1) and the PS1 public science archive have been made possible through contributions by the Institute for Astronomy, the University of Hawaii, the Pan-STARRS Project Office, the Max-Planck Society and its participating institutes, the Max Planck Institute for Astronomy, Heidelberg and the Max Planck Institute for Extraterrestrial Physics, Garching, The Johns Hopkins University, Durham University, the University of Edinburgh, the Queen's University Belfast, the Harvard-Smithsonian Center for Astrophysics, the Las Cumbres Observatory Global Telescope Network Incorporated, the National Central University of Taiwan, the Space Telescope Science Institute, the National Aeronautics and Space Administration under Grant No. NNX08AR22G issued through the Planetary Science Division of the NASA Science Mission Directorate, the National Science Foundation Grant No. AST-1238877, the University of Maryland, Eotvos Lorand University (ELTE), the Los Alamos National Laboratory, and the Gordon and Betty Moore Foundation.

This work is based in part on observations made with the Galaxy Evolution Explorer (GALEX). GALEX is a NASA Small Explorer, whose mission was developed in cooperation with the Centre National d'Etudes Spatiales (CNES) of France and the Korean Ministry of Science and Technology. GALEX is operated for NASA by the California Institute of Technology under NASA contract NAS5-98034.

This research has made use of the NASA/IPAC Extragalactic Database (NED) which is operated by the Jet Propulsion Laboratory, California Institute of Technology, under contract with the National Aeronautics and Space Administration.



\bibliographystyle{mnras}




\bsp	
\label{lastpage}
\end{document}